\documentclass[twocolumn,showpacs,superscriptaddress,preprintnumbers,amsmath,amssymb,prb,aps]{revtex4-1}
\usepackage{graphicx}
\usepackage{dcolumn}
\usepackage{bm}

\begin{document}

\title{Anisotropies in the linear polarization of vacancy photoluminescence in diamond induced by crystal rotations and strong magnetic fields}

\author{D.~Braukmann}
\affiliation{Experimentelle Physik 2, Technische Universit\"at Dortmund, 44227 Dortmund, Germany}

\author{V.~P.~Popov}
\affiliation{Rzhanov Institute of Semiconductor Physics, SB RAS, 630090 Novosibirsk, Russia}

\author{E.~R.~Glaser}
\author{T.~A.~Kennedy}
\affiliation{Naval Research Laboratory, Washington, DC 20375, USA}

\author{M.~Bayer}
\affiliation{Experimentelle Physik 2, Technische Universit\"at Dortmund, 44227 Dortmund, Germany}
\affiliation{Ioffe Institute, Russian Academy of Sciences, 194021 St. Petersburg, Russia}

\author{J.~Debus}
\email[Corresponding author: ]{joerg.debus@tu-dortmund.de}
\affiliation{Experimentelle Physik 2, Technische Universit\"at Dortmund, 44227 Dortmund, Germany}

\begin{abstract}
We study the linear polarization properties of the photoluminescence of the neutral and negatively charged nitrogen vacancies and neutral vacancies in diamond crystals as function of their symmetry and their response to strong external magnetic fields. The linear polarization degree, which exceeds 10\% at room temperature, and rotation of the polarization plane of their zero-phonon lines significantly depend on the crystal rotation around specific axes demonstrating anisotropic angular evolutions. The sign of the polarization plane rotation is changed periodically through the crystal rotation, which indicates a switching between electron excited states of orthogonal linear polarizations. At external magnetic fields of up to 10~T, the angular dependences of the linear polarization degree experience a remarkable phase shift. Moreover, the rotation of the linear polarization plane increases linearly with rising magnetic field at 6~K and room temperature, for the negatively charged nitrogen vacancies, which is attributed to magneto-optical Faraday rotation.
\end{abstract}

\maketitle

\section{Introduction}

Impurities in diamond and silicon carbide have gained remarkable attraction in both fundamental research as well as for technological applications.\cite{MITCF,RTQMW} In diamond the main focus has been drawn on the nitrogen vacancy (NV) center on account of its atom-like properties, namely its discrete level structure.\cite{SiDeCe} Single nanometer-sized diamonds have been grown thus allowing for sensing crystal strain, temperature variations, and electromagnetic fields, which provides, for instance, new possibilities for quantum-based imaging in life sciences.\cite{NaMaSe,temppaper,ElFiSe,McGuinn} The NV center can also serve as a single photon source and material platform to realize high-fidelity quantum entanglement.\cite{QuEnBe,PCMSc}

The NV center consists of a substitutional nitrogen atom paired with a vacancy in the carbon lattice. In case of the negatively charged nitrogen vacancy (NV$^-$) six electrons are localized at the vacancy, whereas the neutral nitrogen vacancy (NV$^0$) has five electrons. For the NV$^-$, the ground and first-excited states are spin-triplet states\cite{ThOfTh}. The optical transition between these states is a zero-phonon line (ZPL) with an energy of 1.946~eV. The ground and excited states of the NV$^0$ are spin-doublet states giving rise to a ZPL at 2.156~eV.\cite{GaliNV0} The basic optical properties of the NV centers have been studied well using, in particular, electron-spin resonance techniques under application of external magnetic fields of a few tens or hundreds of mT.\cite{nphys141,InStMa,neumann,PLDro2} Also, for instance, it has been shown that the excitation of differently orientated single NV centers is polarization dependent.\cite{TPMPRB} Nevertheless, neither the linear polarization properties of the ZPL for an ensemble of NV centers at room temperature nor the impact of large external magnetic fields on the linear polarization characteristics have been investigated. Besides that, the neutral vacancy (V$^0$) in diamond has rarely been characterized by optical methods.

In this paper, we study both the crystallographic as well as magnet-field dependences of the linear polarization properties of the ZPLs of the NV centers and neutral vacancy in diamond crystals. The linear polarization degree and rotation of the polarization plane of the ZPLs of the NV$^-$ and NV$^0$ centers as well as V$^0$ vacancy are significantly changed through crystal rotations around specific axes; their angular dependences are anisotropic. A phenomenological approach based on the symmetry and dipole orientations of the nitrogen-vacancy centers is developed to explain the linear polarization characteristics, in particular for the (001) direction. A strong linear polarization degree exceeding 10\% is found for the NV$^-$ center at room temperature. Moreover, the sign of the polarization plane rotation is altered through the crystal rotation, which demonstrates a switching between radiative electron states of orthogonal linear polarizations. At high external magnetic fields of up to 10~T, the angular dependence of the linear polarization degree in the (001) direction experiences a remarkable phase shift, which may be due to the magnetic torque of the applied field on the magnetic moments of the NV$^-$ center electrons. The rotation of the linear polarization plane, which increases linearly with rising magnetic field at 6~K and room temperature for the NV$^-$ centers, indicates the presence of magneto-optical Faraday rotation.

The paper is organized as follows: In Sec.~\ref{expdetails} the diamond crystals and experimental setup for characterizing the linear polarization degree of the vacancy PL are specified. In Secs.~\ref{angularPLsec} and \ref{bfieldPLsec}, the angular dependence and, respectively, magnetic field dependence of the linear polarization degree and the rotation of the polarization plane are described. The effects of excitation energy and power as well as sample temperature on the linear polarization are presented in Sec.~\ref{tempexcsec}. The paper ends with a conclusion in Sec.~\ref{conclsec}.

\section{Experimental details} \label{expdetails}
\begin{figure}
\centering
\includegraphics[width=8.6cm]{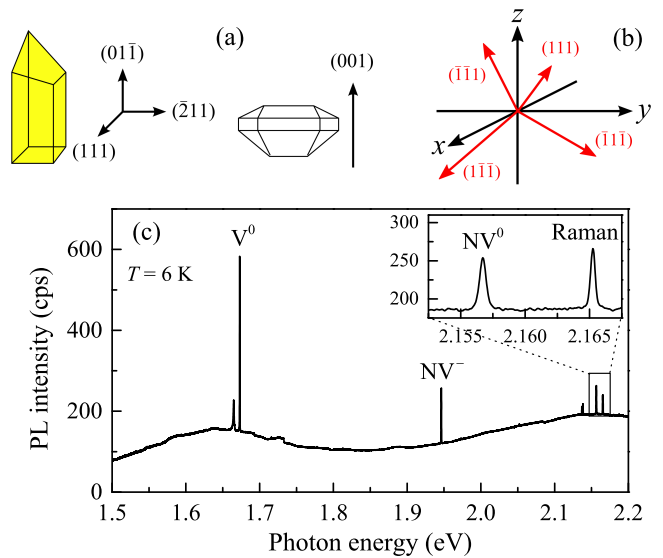}
\caption{\label{fig1} (a) Scheme of the yellow rectangular-prism shaped (left) and transparent rose-cut (right) diamond crystals and their crystallographic symmetry axes that could be addressed optically. (b) Typical orientations of NV centers in a diamond crystal. (c) PL spectrum of the rose-cut diamond crystal excited at 2.33~eV along the $(001)$ direction; $P_{\text{exc}} = 10$~mW, $B=0$~T. Inset: PL spectrum between 2.15 and 2.17~eV shows the ZPL of the NV$^{0}$ centers and the diamond Raman line with 1332~cm$^{-1}$ Raman shift.}
\end{figure}

We studied a transparent rose-cut diamond crystal grown by chemical vapor deposition with an NV center concentration of about 0.14~ppm ($2.5 \times 10^{16}$~cm$^{-3}$) and a yellow rectangular-prism shaped diamond crystal, which had an NV center concentration of about 2.9~ppm ($5.1 \times 10^{17}$~cm$^{-3}$). The latter was grown by high-temperature high-pressure synthesis, electron-irradiated at an energy of 2~MeV and subsequently annealed.\cite{sampleTAK} The rose-cut diamond was optically addressable along the $(001)$ direction, while the latter was excited along the $(111)$ and $(\bar211)$ directions. A scheme of the samples and their symmetry axes is shown in Fig.~\ref{fig1}(a). The samples were mounted strain free on a rotation holder inside a split-coil magnet cryostat, whose magnetic field $\mathbf{B}$ was oriented along the optical axis defined by the direction of the exciting and detected light. The magnetic field was varied between -10 and +10~T and the temperature was changed between 2~K and room temperature. The samples could be rotated around the optical axis with a precision of $\pm 1^\circ$. For excitation, a continuous-wave dye-laser system (Tekhnoscan Ametist-FD-08) equipped with Rhodamine 6G, whose emission energy was tuned between 2.01 and 2.21~eV, and a diode-pumped solid-state laser (Coherent Verdi V10) emitting at 2.33~eV were used. The latter was mainly used to excite the NV$^0$ centers, whereas the NV$^-$ centers and V$^0$ vacancies were excited by either of the laser systems. The exciting laser light was linearly polarized perpendicular to the optical axis. For instance, in consideration of the rectangular-prism shaped diamond crystal and a sample rotation angle $\phi = 0^\circ$ the laser light was linearly polarized along the $(01\bar1)$ direction.

The PL emitted from a diamond crystal was detected with a liquid-nitrogen cooled charge-coupled device camera installed at a single 0.5-m spectrometer. PL pre-studies were also done with a Fourier-transform infrared spectrometer (Bruker Vertex 80v). The PL polarization properties were analyzed by an achromatic half-wave plate and a Glan-Thompson prism. The linear polarization degree of the PL is determined by
\begin{eqnarray}
P_{\text{lin}} = \frac{\sqrt{S_1^2 +S_2^2}}{S_0} 
\label{plins}
\end{eqnarray}
measuring the Stokes parameters $S_0 = I_{0^\circ} + I_{90^\circ}$, $S_1 = I_{0^\circ} - I_{90^\circ}$ and $S_2 = I_{45^\circ} - I_{135^\circ}$ via rotation of the half-wave retarder.\cite{Goldstein} Here, the index of the PL intensity $I$ indicates the angle of the linear polarization plane of the PL with respect to the linear polarization plane of the exciting laser light, which was not changed throughout the measurements. By analogy with the Faraday rotation angle, we define the rotation of the linear polarization plane by
\begin{eqnarray}
\theta = \frac{180^{\circ}}{2\pi} \text{arctan}\left(\frac{S_2}{S_1}\right) .
\end{eqnarray}
For analyzing the angular dependence of the linear polarization degree, $\theta$ can be seen as the angular change of $P_{\text{lin}}$.

\section{Angular dependence of linear polarization of PL} \label{angularPLsec}

\begin{figure*}
\centering
\includegraphics[width=17.8cm]{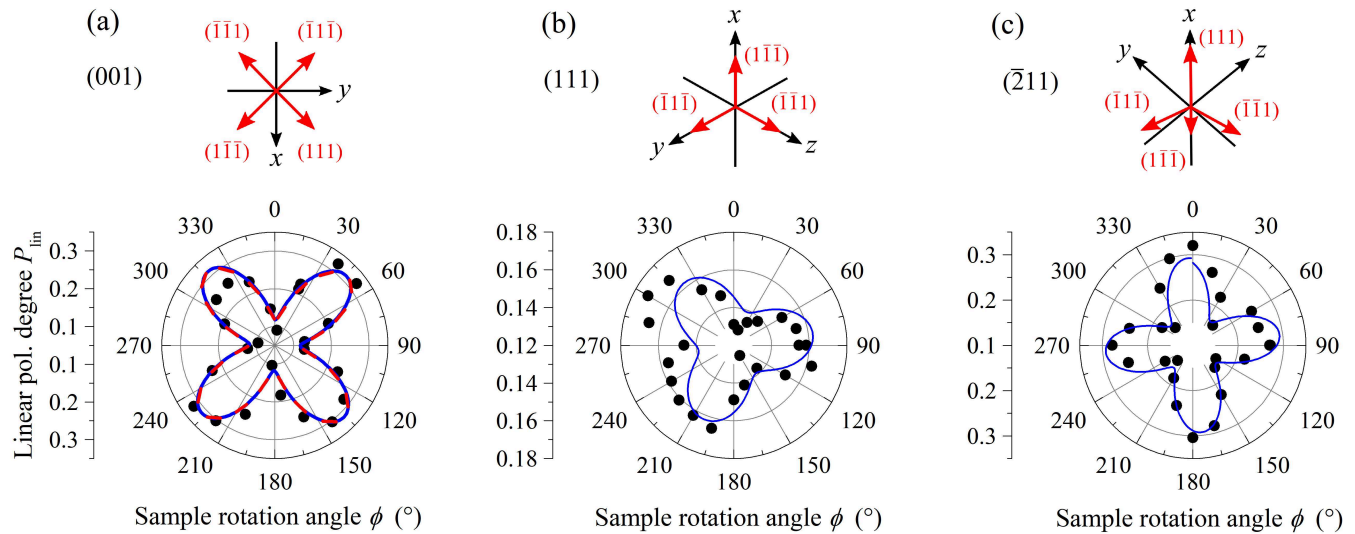}
\caption{\label{fig2} For the optical axis along the (a) $(001)$, (b) $(111)$, and (c) $(\bar211)$ direction, the 2D projection of the NV center orientations (red arrows) and the dependence of the linear polarization degree (black dots) of the NV$^-$ ZPL on the sample rotation angle are shown; $T = 6$~K, $B = 0$~T, $E_{\text{exc}} = 2.33$~eV for $(001)$ and 2.21~eV for $(111)$ and $(\bar211)$. The blue solid lines are sine-based fit functions according to Eq.~\eqref{sinefit}, the red dashed line in (a) is a fit function based on Eq.~\eqref{eqpiphi2}. The experimental error does not exceed the symbol size.}
\end{figure*}

The influence of the diamond-crystal and NV center axes orientations on the linear polarization properties of the vacancy PL will be investigated in the following. A spectrally broad-ranged PL spectrum of the rose-cut diamond crystal is shown in Fig.~\ref{fig1}(c); the intensity $I_{0^\circ}$ was measured. It includes the zero-phonon lines of the neutral (2.157~eV) and negatively charged (1.947~eV) NV centers, neutral vacancies V$^0$ (1.674~eV and 1.666~eV), their respective phonon sidebands, and the diamond Raman line at 2.165~eV, for an optical excitation at 2.33~eV. We will focus on the V$^{0}$ vacancies and, in particular, on the NV$^{0}$ and NV$^{-}$ centers.

According to the lattice symmetry of diamond NV centers are oriented along the $(111)$, $(1\bar1\bar1)$, $(\bar11\bar1)$ and $(\bar1\bar11)$ directions.\cite{InStMa} These possible NV center orientations, which are described by the C$_{3\text{v}}$ symmetry group, are depicted in Fig.~\ref{fig1}(b). For our experiments, the two-dimensional (2D) projections of the NV center orientations along the optically accessible crystallographic axes are relevant. In Fig.~\ref{fig2}(a) the 2D projection along the $(001)$ direction is sketched; one can clearly see that the three-dimensional arrangement of the NV center axes according to a tetrahedral symmetry is reduced to a rectangular cross (see red arrows) lying within the $xy$-plane. Below this sketch in Fig.~\ref{fig2}(a) the dependence of $P_{\text{lin}}$ of the NV$^-$ ZPL on the diamond crystal rotation around the $(001)$ direction is demonstrated. $P_{\text{lin}}$ varies between 0.09 and 0.32 with a 90$^\circ$ periodicity, which follows from a sine-function fit (see blue solid line) that is given by
\begin{eqnarray}
P_{\mathrm{lin}} = K \text{sin} \left ( \frac{2\pi}{p} \phi +\phi_0 \right ) + \kappa .
\label{sinefit}
\end{eqnarray}
Here, the amplitude $K = 0.098 \pm 0.008$, periodicity $p = (90.6 \pm 0.9)^{\circ}$, phase shift $\phi_0 = (0 \pm 2)^{\circ}$ and offset $\kappa = 0.215 \pm 0.005$ are four independent fitting parameters. The maxima of $P_{\text{lin}}$ are at about $\phi = 45^\circ$, $135^\circ$, $225^\circ$ and $315^\circ$. At these angles the linear polarization plane, namely the electric field vector of the NV$^{-}$ ZPL emission, coincides with an NV$^{-}$ center symmetry axis. The minima are observed at angles, at which the linear polarization plane lies between two of the NV-center symmetry axes. Apparently, this anisotropic angular dependence of $P_{\text{lin}}$ mirrors the rectangular symmetry of the 2D projections of the NV center orientations.

The angle-dependent $P_{\text{lin}}$ of the NV$^-$ ZPL emitted along the $(111)$ direction is shown in Fig.~\ref{fig2}(b). The overall change in the polarization degree only amounts to about $4$~\%. The maxima and minima vary with a 120$^\circ$ periodicity. In fact this periodicity is attributed to the tripod symmetry of the corresponding 2D projection of the NV center orientations, see sketch at the top of Fig.~\ref{fig2}(b). The tripod symmetry appears because the NV centers, which are oriented along the $(111)$ axis, coincide with the optical axis. For the NV$^-$ ZPL emitted along the $(\bar211)$ direction, $P_{\text{lin}}$ is shown as function of $\phi$ in Fig.~\ref{fig2}(c). Here, $P_{\text{lin}}$ varies between 0.15 and 0.32 with a 90$^\circ$ periodicity, which is similar to that shown in Fig.~\ref{fig2}(a). The 2D projection of the NV center orientations demonstrates a reflection symmetry along the $x$-axis. Moreover, the projections of the NV-center symmetry axes differ in their lengths. For example, the $(1\bar1\bar1)$ axis is almost parallel to the optical axis; therefore, its projection vector is rather short. Thus, one cannot actually expect a 90$^\circ$ periodicity. Only the maximum of $P_{\text{lin}}$ at about 0$^\circ$, which coincides with the $(111)$ axis, supports a relation between the angular $P_{\text{lin}}$-dependence and the 2D projection. Such a simple relation cannot be found for the other maxima of $P_{\text{lin}}$ and the corresponding NV center directions.

\begin{figure*}
\centering
\includegraphics[width=17.8cm]{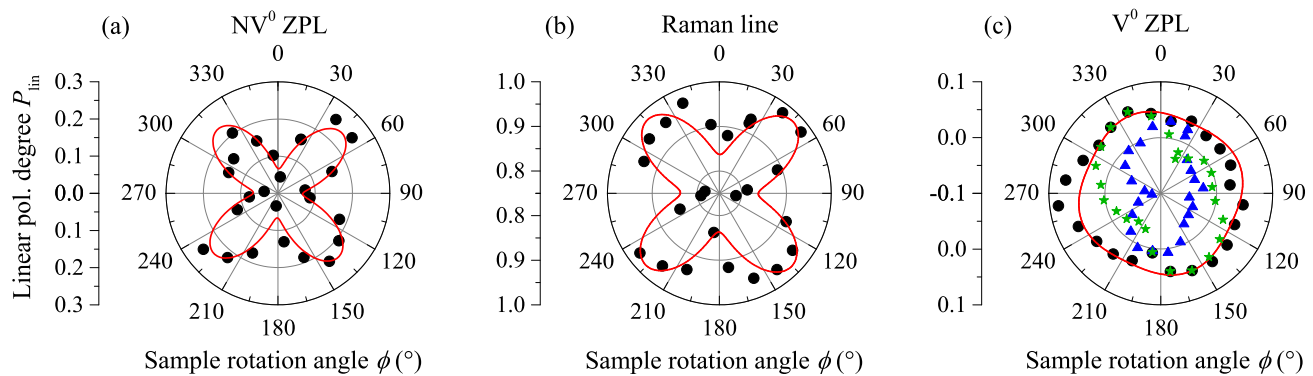}
\caption{\label{fig3} Dependence of the linear polarization degree $P_{\text{lin}}$ (black dots) of the (a) NV$^0$ ZPL, (b) Raman line, and (c) V$^0$ ZPL on the sample rotation angle $\phi$; $T = 6$~K, $B = 0$~T, $E_{\text{exc}} = 2.33$~eV. The optical axis is oriented along the $(001)$ direction. In (c) the Stokes parameters $S_1/S_0$ (blue triangles) and $S_2/S_0$ (green stars) are also shown. The fit curves (red solid lines) are based on Eq.~\eqref{eqpiphi2}. The experimental error does not exceed the symbol size.}
\end{figure*}

Now, we like to evaluate a relation between the linear polarization degree of the exciting laser light and that of the NV$^-$ PL by considering the diamond crystal symmetry and NV center orientations. Our main focus will be drawn on the high-symmetric $(001)$ direction. The components of the electric field vectors of the PL, $E_{i,j}^{\mathrm{PL}}$, and exciting light, $E_{k,n}^{\mathrm{exc}}$, are linked through the rank-four tensor $A_{ijkn}$ as follows:
\begin{eqnarray}
\langle E_i^{\mathrm{PL}} E_j^{\mathrm{PL} \ast} \rangle = A_{ijkn} \langle E_k^{\mathrm{exc}} E_n^{\mathrm{exc} \ast} \rangle .
\end{eqnarray}
Here, $\langle E_i^{\mathrm{PL}} E_j^{\mathrm{PL} \ast} \rangle$ and $\langle E_k^{\mathrm{exc}} E_n^{\mathrm{exc} \ast} \rangle$ are the components of the non-normalized polarization tensor.\cite{LDLTCT} In our phenomenological approach, $A_{ijkn}$ shall only be determined by the orientation of the NV center in diamond. For an optical excitation along the $(001)$ direction, the 2D projection of the NV center axes lies within the $xy$-plane and is shaped like a rectangular cross, see Fig.~\ref{fig2}(a). Accordingly, the three-dimensional symmetry of the four NV centers is reduced to a cubic symmetry. For the cubic symmetry, one obtains $A_{iiii} = A_{11}$,  $A_{iijj} = A_{12}$,  $A_{ijij} = A_{44}$ and $A_{ijji} = A_{47}$ with $i \neq j$, while the other tensor components are zero.\cite{AotMAA} Since only the linear polarization is considered and the paths of the exciting laser beam and PL emission are parallel to the $(001)$ axis, the linear polarization degree can be expressed by
\begin{eqnarray}
P_{\mathrm{lin}}(\phi) &=& \frac{b}{2a} \cos(4 \phi +\phi_0) + \frac{b+2c}{2a}  \nonumber \\
 &=& A \cos(4 \phi +\phi_0) + d . \label{eqpiphi2}
\end{eqnarray}
The parameters are given by $a=A_{11} +A_{12}$, $c= A_{44}+A_{47}$ and $b=A_{11}-A_{12}-A_{44}-A_{47}$, which are summarized by $A = b/2a$ and $d =(b+2c)/2a$. Additionally, we deduce the relation for the angular dependence of the rotation of the linear polarization plane from the fact that $\theta$ can be understood as the angular change of $P_{\mathrm{lin}}$: 
\begin{eqnarray}
\frac{\text{d} P_{\mathrm{lin}}(\phi)}{\text{d}\phi} \propto \theta (\phi) = C \cos(4 \phi +\phi_0 + 45^\circ) + e
\label{eqtephi}
\end{eqnarray}
with some amplitude $C$ and offset $e$. The angular dependence of $P_{\mathrm{lin}}$ can be well modeled by Eq.~\eqref{eqpiphi2}, as is shown by the red dashed line in Fig.~\ref{fig2}(a). The fitting parameters $A = 0.098 \pm 0.008$, $\phi_0 = (-1 \pm 1)^{\circ}$ and $d = 0.216 \pm 0.005$ are very similar to the parameters obtained by the sine-based fit function (blue solid line) according to Eq.~\eqref{sinefit}.

For gaining further insight into the anisotropic angular evolution of $P_{\text{lin}}$, we take account of the fine structure of the NV$^-$ centers. Their ZPLs are spin-conserving optical transitions between the excited $^3$E and ground $^3$A$_2$ states. Since we study the linear polarization properties of the PL, we focus on the states that are characterized by zero spin and orbital angular momentum projections. The photons, emitted during the electron relaxation process, are polarized along the dipole axis, which is perpendicular to the symmetry axis of the NV centers.\cite{nphys141} Vice versa, in order to excite an electron to linearly polarized states the polarization of the laser light must also be perpendicular to the NV center symmetry axis, in particular, for resonant excitation.\cite{QuEnBe}

In the case of optical excitation along the $z$-axis, the exciting light is polarized, for instance, along the $xy$-diagonal, compare Fig.~\ref{fig2}(a); thus, only two NV$^-$ centers are linearly excited. In turn, solely these two NV$^-$ centers contribute to the emission process of the linearly polarized ZPL. Since the dipoles of these centers are oriented parallel, the polarization plane of the ZPL emission from these NV$^-$ centers are also parallel; hence, $P_{\text{lin}}$ is maximized. However, when the exciting light is polarized along the $x$- or $y$-axis, every NV$^-$ center is excited and contributes to the ZPL emission process. As the dipole orientations of the $(\bar1\bar11)$ and $(111)$ NV$^-$ centers are perpendicular to the orientations of the $(\bar11\bar1)$ and $(1\bar1\bar1)$ NV$^-$ center dipoles, the polarization planes of the ZPL emission from these centers are perpendicular, too. Therefore, the total linear polarization degree is reduced. Note that a linearly polarized excitation of the NV$^-$ centers in fact creates electrons in both linearly and circularly polarized excited states. As a consequence, the ZPL emission consists of linearly and also circularly polarized contributions. A detailed analysis of the circular polarization properties of the NV center PL is reported in Ref.~\onlinecite{JDdiaCP}.

\begin{figure*}
\centering
\includegraphics[width=17.8cm]{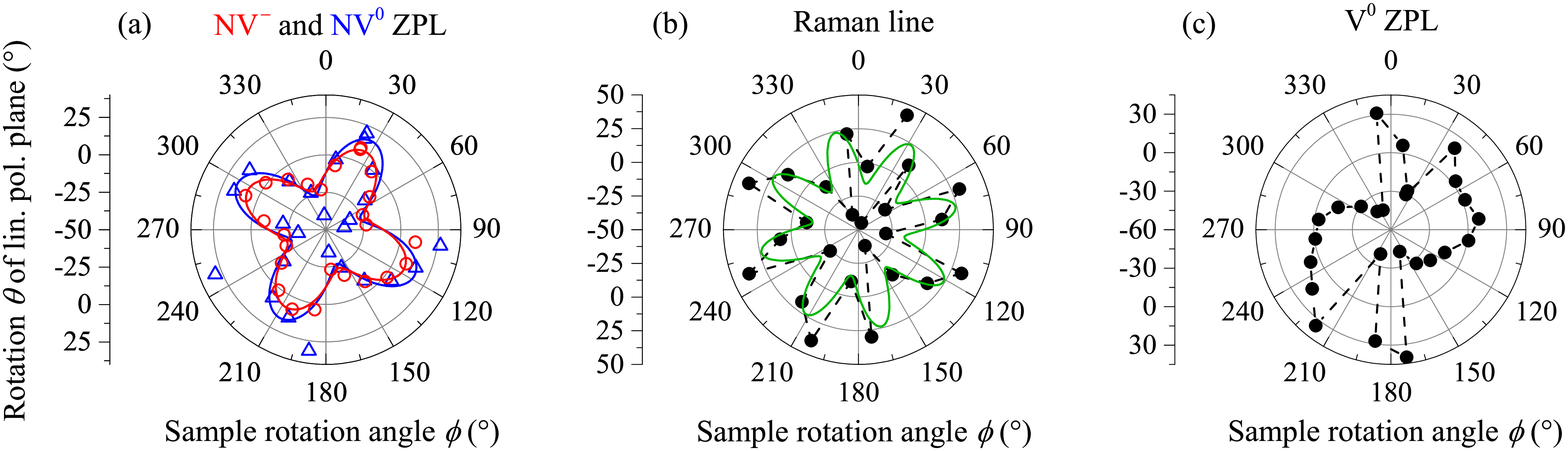}
\caption{\label{fig4} Rotation of the plane of the linear polarization degree, for (a) the NV$^-$ (circles) and NV$^0$ (triangles) ZPLs, (b) Raman line, and (c) V$^0$ ZPL, as function of the crystal rotation angle; $T = 6$~K, $B = 0$~T, $E_{\text{exc}} = 2.33$~eV. The optical axis is oriented along the $(001)$ direction. The red and blue-colored curves are fit functions based on Eq.~\eqref{eqtephi}, the green-colored curve is a sine-function fit. The experimental error does not exceed the symbol size. In (b) and (c) the dashed lines are guides to the eye.}
\end{figure*}

In the following we like to extend our phenomenological approach to the linear polarization characteristics of other vacancies in the diamond crystals studied and their Raman line observed at 1332~cm$^{-1}$. In Fig.~\ref{fig3}(a) $P_{\text{lin}}$ of the NV$^0$ center ZPL is depicted as a function of the sample rotation angle. Since the symmetry of the NV$^-$ centers coincides with that of the NV$^0$ centers, we also see an anisotropic angle dependence of $P_{\text{lin}}$ with a 90$^\circ$ periodicity. Here, $P_{\text{lin}}$ varies between 0.02 and 0.25; in absolute terms it is about 0.07 smaller than the values for the NV$^-$ ZPL. The angular evolution can be described suitably by Eq.~\eqref{eqpiphi2}.

The linear polarization properties of the 1332~cm$^{-1}$ Raman line, which indicates a vibration of the two diamond Bravais lattices relative to each other,\cite{RaDi45,RaDi70} are shown in Fig.~\ref{fig3}(b). The Raman line is strongly linearly polarized, $P_{\text{lin}}$ changes between 0.97 and 0.78. As in the case of the NV centers, the angular evolution follows Eq.~\eqref{eqpiphi2}. Hence, our phenomenological approach is valid not only for the PL, but also for inelastic light scattering. This agrees well with a study of Mossbrucker et al.~\cite{JMDol} reporting on a similar anisotropy in the polarization degree of the Raman line in diamond along the $(001)$ direction. They demonstrated a maximal Raman-line polarization, as in our case, when the polarization of the incident light was parallel to the 2D projections of the diamond lattice and the optical axis. In this configuration the polarizability of the diamond crystal is maximal; therefore, the polarization of the photons is preserved during the Raman scattering process and the polarization degree is close to unity.

For the neutral vacancy V$^0$, which consists of a single vacancy in the diamond lattice, $P_{\text{lin}}(\phi)$ of its ZPL at 1.674~eV is presented in Fig.~\ref{fig3}(c). Note that the weak ZPL at 1.666~eV will not be discussed in the following. It exhibits small values varying between 0.02 and 0.08. Eq.~\eqref{eqpiphi2} only approximately describes the data, see the red solid curve. The symmetry considerations can be done by analogy with that of the NV centers, since the 2D projection of the neutral vacancies having T$_{\text{d}}$ symmetry in the (001) direction is comparable with the projection of the NV centers of C$_{\text{3v}}$ symmetry.\cite{Ocosii,V0MRM} In contrast to the NV centers, the V$^0$ does not possess a static dipole moment due to the symmetric electron distribution. As a result, $P_{\text{lin}}$ is small and a significant anisotropy in its angle dependence cannot be observed. Only the normalized Stokes parameters $S_1$ and $S_2$ show a 180$^{\circ}$ anisotropy with symmetry axes around 0$^{\circ}$, for $S_1/S_0$ (blue triangles), and 135$^{\circ}$, for $S_2/S_0$ (green stars), see Fig.~\ref{fig3}(c). This anisotropy may arise from the static Jahn-Teller distortion which reduces the T$_{\text{d}}$ to a D$_{\text{2d}}$ symmetry. However, theoretical calculations have shown that the dynamic Jahn-Teller distortion, which causes the T$_{\text{d}}$ symmetry, is much more likely.\cite{FPSJT}

In addition to the linear polarization degree, the rotation $\theta$ of the linear polarization plane is a further measure of the polarization properties of the vacancy PL and crystal symmetry. For the NV$^-$ and NV$^0$ ZPLs, one obtains 90$^\circ$-periodic dependences of $\theta$ on the crystal rotation angle around the $(001)$ axis, as shown in Fig.~\ref{fig4}(a). They are in agreement with Eq.~\eqref{eqtephi}. Besides that, the sign and absolute values of $\theta$ depend significantly on $\phi$; $\theta$ changes between $+8^\circ$ and $-23^\circ$, for NV$^-$, and $+32^\circ$ and $-40^\circ$, for NV$^0$. The sign changes in the rotation of the $P_{\text{lin}}$-plane indicate a switching between radiative electron states of orthogonal linear polarizations. One may describe the rotations of the diamond crystal by $45^\circ$ as a binary polarization modulator, which selects the contributions of the orthogonal transition dipole\cite{nphys141} states to the linearly polarized ZPL emission.

Similarly strong changes in the minimum and maximum values of $\theta$ are found for the Raman scattering line, see Fig.~\ref{fig4}(b), and for the V$^0$ ZPL, as presented in Fig.~\ref{fig4}(c). They vary between about $-40^\circ$ and $+40^\circ$. In contrast to the 90$^\circ$ periodicity observed for the NV centers and the angular dependence of $P_{\text{lin}}$ of the Raman line, $\theta(\phi)$ of the Raman line demonstrates a 45$^\circ$ periodicity, which can be fit best with a sine-function. The dependence $\theta(\phi)$ of the neutral vacancies is 180$^\circ$ periodic and cannot be described by a simple sine function. Apparently, the phenomenological approach to deduce an equation for $\theta$ is specifically valid for NV centers rotated around the $(001)$ crystal axis, where the assumption of a cubic symmetry applies.

\section{Magnetic-field dependent linear polarization of the NV$^-$ ZPL} \label{bfieldPLsec}

\begin{figure*}
\centering
\includegraphics[width=17.8cm]{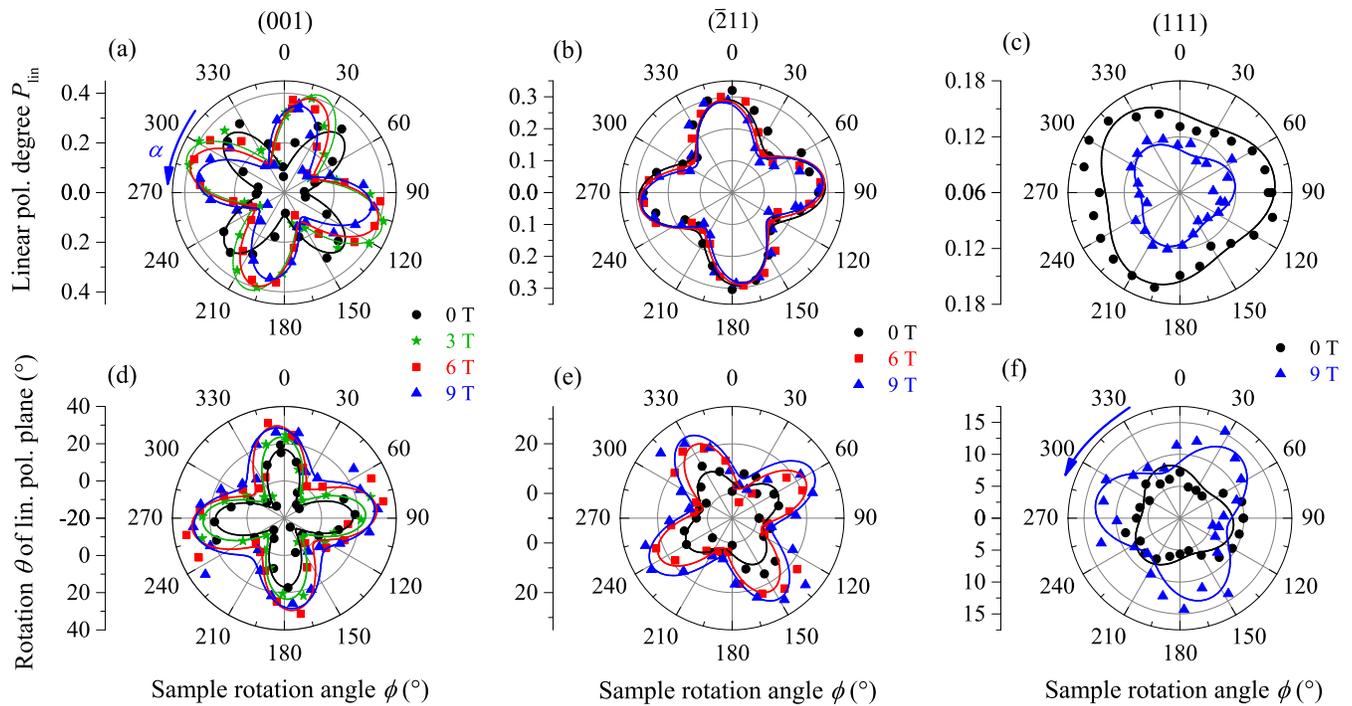}
\caption{\label{fig7} Magnetic field dependence of the linear polarization degree and rotation of the polarization plane of the NV$^-$ ZPL along the (a,d) $(001)$, (b,e) $(\bar211)$ and (c,f) $(111)$ direction as function of the sample rotation angle; $T = 6$~K, $E_{\text{exc}} = 2.21$~eV. The curves are fits based on Eqs.~\eqref{eqtephi} and \eqref{eqpiphi3} or in case of the $(111)$ direction a function $\propto \text{sin}(3 \phi)$. The experimental error does not exceed the symbol size.}
\end{figure*}

In the following the impact of an external magnetic field up to 10~T on the linear polarization properties of the NV$^-$ ZPL will be described. In Fig.~\ref{fig7}(a) the linear polarization degree is shown as function of the crystal rotation angle around the $(001)$ direction for four different magnetic field strengths. The absolute values of $P_{\text{lin}}$ do not vary significantly with increasing magnetic field strength. This is due to the fact that the linearly polarized excited states with zero spin involved in the radiative electron relaxation process differ in their orbital angular momentum only. The respective orbital Zeeman splitting is weak, since the g-factor is only about $0.1$,\cite{gfactorpaper} and thus strong changes in the population of the states cannot be introduced with magnetic fields up to 9~T. Moreover, $P_{\text{lin}}$ shows similar 90$^\circ$ periodic angle dependences at magnetic fields between 0 and 9~T. However, the dependences demonstrate a remarkable phase shift: At zero magnetic field $P_{\text{lin}}$ has its maximum at about $\phi = 45^\circ$, while with increasing magnetic field strength it gradually shifts to lower angles reaching 11$^\circ$ at 9~T. This magnetic-field dependent phase shift $\alpha$ shall be introduced in the mathematical description of the anisotropy of the linear polarization degree. Accordingly, Eq.~\eqref{eqpiphi2} is modified in the following way:
\begin{eqnarray}
P_{\text{lin}}(\phi) = A \cos \left [ 4 \phi +\phi_0 +\alpha(B) \right ] + d .
\label{eqpiphi3}
\end{eqnarray}
This equation is used to fit the data shown in Fig.~\ref{fig7}(a), see solid curves.

The phase shift may result from the magnetic torque of the external field acting on the magnetic moments of the NV$^-$ center electrons,\cite{Rand,Hansen} so that they tend to align in such a way that the total magnetic energy is minimized. Such a rotation of the magnetic moments could lead to a rotation of the angular dependence of $P_{\text{lin}}$, since the polarization plane of the emitted ZPL photons is perpendicular to the dipole orientation. Alternatively, one may assume that the phase shift observed is due to a mixing of states with different magnetic moments under the application of a magnetic field specifically oriented with respect to the NV$^-$ center symmetry axes.

This phase shift of $P_{\text{lin}}$ can only be seen in the $(001)$ direction. As depicted in Fig.~\ref{fig7}(b), there is no significant phase shift in the $(\bar211)$ direction. The angular dependence takes almost the same form, and a variation in the $P_{\text{lin}}$ amplitude cannot be observed. In Fig.~\ref{fig7}(c) the angle dependence of $P_{\text{lin}}$ along the $(111)$ direction is shown. Here, a significant phase shift of $P_{\text{lin}}$ with rising magnetic field is not detected, but $P_{\text{lin}}$ increases slightly by about 0.03 from 0 to 9~T. Note that, for the $(001)$ and $(\bar211)$ directions, this increase is in the order of the experimental error; therefore, one cannot certainly conclude that such an increase is not present in these directions.

\begin{figure}
\centering
\includegraphics[width=8.6cm]{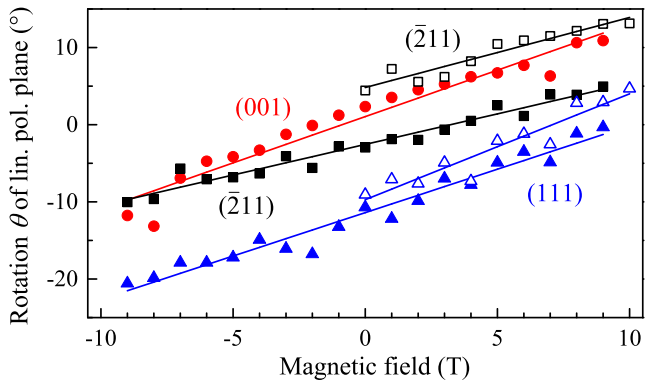}
\caption{\label{fig8} Magnetic field dependence of $\theta$ for different crystal directions measured at 6~K (closed symbols) and room temperature (open symbols). The solid lines are fits following Eq.~\eqref{eqfit}; $E_{\text{exc}} = 2.21$~eV. The experimental error does not exceed the symbol size.}
\end{figure}

In contrast to the angular dependence of $P_{\text{lin}}$, the polarization-plane rotation angle $\theta$ does not show a significant phase shift depending on the magnetic field strength, for the optical axis oriented along the $(001)$ direction, as illustrated in Fig.~\ref{fig7}(d). However, we observe a phase shift in $\theta$, for the optical excitation along the $(111)$ direction. As one can see in Fig.~\ref{fig7}(f), the angular shape of $\theta$ is rotated by about 60$^{\circ}$. This phase shift is larger than that of $P_{\text{lin}}$ in the $(001)$ crystal direction. Besides that, the value of the 90$^\circ$-periodic oscillation, for the (001) direction, changes drastically: While at zero magnetic field $\theta$ varies between $-14^\circ$ and $+19^\circ$, at 9~T the variation lies between $+2^\circ$ and $+32^\circ$. In Fig.~\ref{fig8} the dependence of $ \theta$ on $B$ is shown for the $(001)$ direction measured at $T = 6$~K (red dots). The angle $\phi$ was set to a value of maximum linear polarization degree. One clearly sees that $\theta$ increases linearly with rising magnetic field. Accordingly, the description of the anisotropic behavior of $\theta(\phi)$ shall be extended through a variable offset $e$ given previously in Eq.~\eqref{eqtephi}:
\begin{eqnarray}
e = e_0 + e(B) \cdot B 
\label{eqfit}
\end{eqnarray}
with $e_0 = (1.0 \pm 0.4)^{\circ}$ and $e(B) = (1.19 \pm 0.07)^{\circ}\text{/T}$, for the $(001)$ direction. It is worthwhile to mention that the Faraday rotation of the cryostat windows, which also caused an increase in $\theta$ with rising magnetic field, was separately measured and was corrected in Eq.~\eqref{eqfit}. Moreover, we will show in the following that the linear increase of $\theta(B)$ is observed for all crystal directions studied.

In Fig.~\ref{fig7}(e) the angular dependence of $\theta$ in the $(\bar211)$ direction is depicted. As in the case of the $(001)$ direction a phase shift cannot be observed, but $\theta$ increases linearly with rising magnetic field, see also the solid squares in Fig.~\ref{fig8}. The slope is evaluated to $e(B) = (0.79 \pm 0.04)^{\circ}\text{/T}$. For the $(111)$ direction, the $\theta$-dependences are shown in Fig.~\ref{fig7}(f) and Fig.~\ref{fig8} (see triangles). According to Eq.~\eqref{eqfit} we obtain $e_0 = (-11.4 \pm 0.3)^{\circ}$ and $e(B) = (1.13 \pm 0.05)^{\circ}\text{/T}$, which coincides within the limits of accuracy with the $e(B)$ obtained for the $(001)$ direction. Apparently, this increase seems to only weakly depend on the crystal orientation. The arithmetic mean value of the magnetic-field dependent increase of $\theta$ is given by $(1.04 \pm 0.09)^{\circ}\text{/T}$. Moreover, $\theta(B)$ shows a similar behavior at room temperature, as demonstrated by the open symbols in Fig.~\ref{fig8}. In (111) direction, $e(B) = (1.4 \pm 0.2)^{\circ}\text{/T}$ and, in ($\bar2$11) direction, $e(B) = (0.91\pm0.09)^{\circ}\text{/T}$ are obtained. These values are within the margin of the error in line with the values evaluated at $T=6$~K. Thus, the anisotropy of $\theta$ and its magnetic field dependence are quite stable against temperature changes. Only the error is increased due to a small ZPL intensity at high temperature.\cite{SiDeCe} Note that, for the ZPL at 1.19~eV resulting from the singlet state transition,\cite{InfraRod} $P_{\text{lin}}$ amounts to about 0.15 irrespective of the magnetic field strength, and $\theta$ rises with $(2.0 \pm 0.4)^{\circ}\text{/T}$ measured along the (111) direction.

The magnetic field dependence of $P_{\text{lin}}$ highlights that not only crystallographic properties are relevant for the linear polarization properties of the NV$^-$ centers in diamond. The rotation of the linear polarization plane of the NV$^-$ center emission with respect to that of the incident laser light also indicates the presence of magneto-optical rotation: The ensemble of NV centers in the diamond crystals having ground and excited states subjected to an external magnetic field exhibits birefringence in its response to the linearly polarized optical laser field. This originates from the asymmetry in the refractive indices for the left- and right-circularly polarized light caused by the Zeeman splitting of the magnetic sublevels.\cite{Patnaik,Savchuk} Accordingly, one could refer to the parameter $e(B)$ as the Verdet constant $V$. However, our results, for the (001) and (111) or ($\bar2$11) directions, do not depend on the thickness $\eta$ of the crystal, which one would expect according to the equation $V = \theta/B\eta$ characterizing the magneto-optical (Faraday) rotation. It is quite probable that mainly the NV$^-$ centers at the surfaces of the diamond crystals are excited and contribute to the emission, so that the area of optically active NV$^-$ centers is similar in both crystals studied. The magnetic-field induced circular dichroism in diamond itself is negligible, since the absorption edge of diamond is in the deep ultra-violet spectral range.

\section{Dependence of the linear polarization on the temperature, laser excitation energy and power} \label{tempexcsec}

\begin{figure}
\centering
\includegraphics[width=8.6cm]{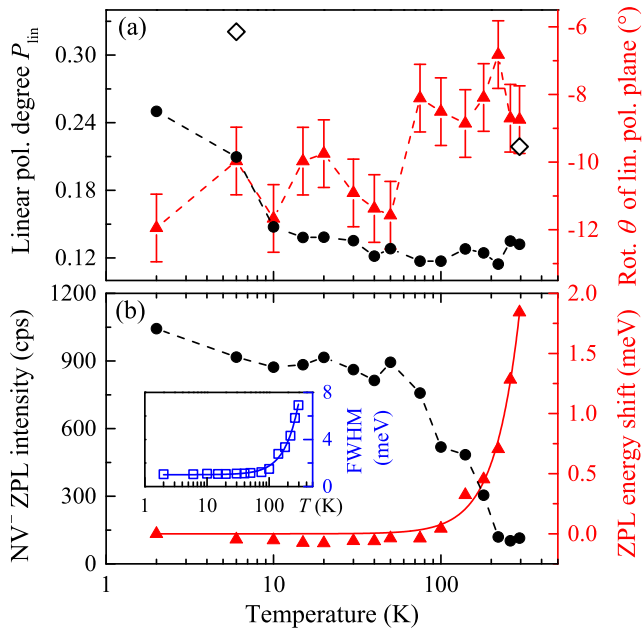}
\caption{\label{fig6} (a) Temperature dependence of the linear polarization degree (left scale) and rotation of the linear polarization plane (right scale) of the NV$^-$ ZPL; $E_{\text{exc}} = 2.21$~eV, $P_{\text{exc}} = 10$~mW. (b) Intensity (left scale) and energy (right scale) of the NV$^-$ ZPL as a function of the temperature. The optical axis is oriented along the $(111)$ direction. The fit function (solid curve) is a power function. Dashed lines are guides to the eye. Inset: Temperature dependence of the full width at half maximum of the NV$^-$ ZPL. Unless shown otherwise, the experimental error does not exceed the symbol size.}
\end{figure}

Finally, the impacts of the energy and power of the exciting laser light as well as the sample temperature on the linear polarization characteristics of the NV$^-$ center emission were also studied. The temperature measurements were performed with an excitation along the $(111)$ direction. A sample rotation angle of $\phi = 90^\circ$ was chosen to have maximum $P_{\text{lin}}$. Fig.~\ref{fig6}(a) shows the temperature dependence of $P_{\text{lin}}$ (black dots) and $\theta$ (red triangles) of the NV$^-$ ZPL. At a temperature of about 2~K $P_{\text{lin}}$ reaches 0.25, it however decreases rapidly for increasing the temperature up to about 15~K. Here, it reaches a value of about $0.13 \pm 0.01$. It does not decrease further; at room temperature the linear polarization degree remains at about $13\%$. In contrast, $\theta$ demonstrates a weak temperature dependence. At temperatures below 50~K $\theta$ is around $(-11 \pm 2)^{\circ}$, and above 50~K it slightly increases to about $(-8 \pm 2)^{\circ}$. It is worthwhile to mention that the linear polarization degree takes higher values in other geometries: When the optical axis is directed along the $(\bar211)$ direction, $P_{\text{lin}}$ is about 0.32 at 6~K and 0.22 at room temperature, see open diamonds in Fig.~\ref{fig6}(a). Thus, the linear polarization degree of the NV$^-$ ensemble ZPL demonstrates a quite high temperature robustness.

For a single NV$^-$ center it was found that at temperatures above 40~K all polarization of the ZPL emission is vanished due to a population transfer between the excited states of the NV$^-$ center.\cite{FuJTE} This population transfer is coupled to a degenerated vibrational mode, which is a consequence of the Jahn-Teller effect.\cite{GaliJTE, AbtewJTE} In addition to the decrease of polarization, another consequence of this population transfer is a strong broadening of the NV$^-$ ZPL following a $T^{5}$-dependence. In comparison to that, we observe a weaker temperature dependence. As illustrated in the inset of Fig.~\ref{fig6}(b), the full width at half maximum (FWHM) of the NV$^-$ ZPL shows a $T^{1.9 \pm 0.1}$-dependence. Moreover, the ZPL energy decreases by following a $T^{2.9 \pm 0.2}$-dependence, see red triangles in Fig.~\ref{fig6}(b). This behavior can be related to a direct phonon process between excited states; Raman processes can be neglected, since they typically show temperature dependences of higher order. Furthermore, the ZPL intensity reduces with increasing temperature; its evolution can be described by the Debye-Waller factor $e^{-\coth(1/T)}$. These temperature dependent characteristics of the ZPL can more likely be attributed to an electron-phonon interaction described by the Franck-Condon-coupling.\cite{FrCoKKR} It seems, also due to the quite high $P_{\text{lin}}$ at room temperature, that the Jahn-Teller effect is less important for an ensemble of NV$^-$ centers.

\begin{figure}
\centering
\includegraphics[width=8.6cm]{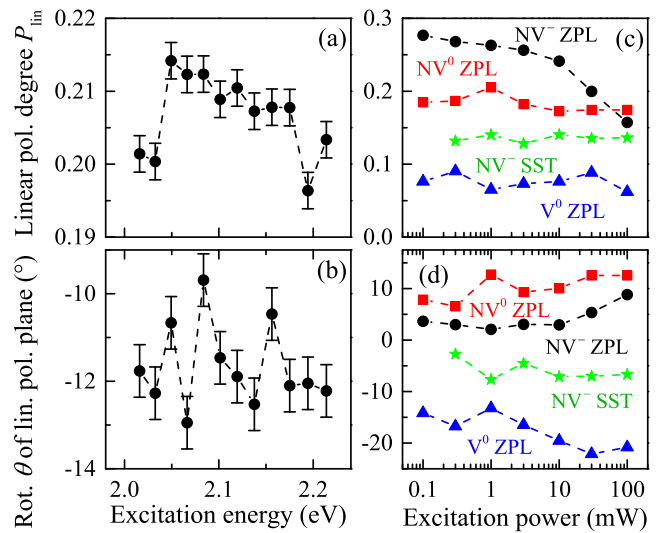}
\caption{\label{fig5} Dependence of (a) $P_{\text{lin}}$ and (b) $\theta$ of the NV$^-$ ZPL on the laser excitation energy; $T=6$~K, $B=0$~T, $P_{\text{exc}} = 3$~mW. Dashed lines are guides to the eye. (c) $P_{\text{lin}}$ and (d) $\theta$ of the NV$^-$ ZPL (black dots), NV$^0$ ZPL (red squares), V$^0$ (blue triangles) and the singlet state transition of the NV$^-$ centers (green stars) as function of the laser excitation power; optical axis is oriented along the $(001)$ direction and $E_{\text{exc}} = 2.21$~eV. Unless shown otherwise, the experimental error does not exceed the symbol size.}
\end{figure}

In Figs.~\ref{fig5}(a) and \ref{fig5}(b) the dependences of $P_{\text{lin}}$ and $\theta$ on the excitation energy are shown. A variation of the laser excitation energy from 2.01 to 2.21~eV leads neither to a significant change in the linear polarization degree nor to an alteration in the polarization plane rotation. Within the margin of error, $P_{\text{lin}}$ is about constant, slightly fluctuating around $0.21 \pm 0.01$, and $\theta$ is approximately $(11 \pm 2)^\circ$. Apparently, the excitation energy is of minor relevance in the region of highest absorption for the NV$^-$ centers.\cite{MitaCabsor} An enhancement of $P_{\text{lin}}$ for quasi-resonant excitation of the NV$^-$ centers is not observed.

In Fig.~\ref{fig5}(c) the laser power dependence of $P_{\text{lin}}$ for the NV$^-$ center emission is depicted (black dots). For a low laser power, $P_{\text{lin}}$ stays constant at a value of 0.27, but when the power increases above about 3~mW, $P_{\text{lin}}$ decreases down to 0.16. This is however not the case for the $P_{\text{lin}}$ of the NV$^0$ (red squares) and V$^0$ (blue triangles) ZPLs. The linear polarization degree of the singlet-state transition (SST) PL\cite{InfraRod} of the NV$^-$ centers at 1.19~eV (green stars) remains also stable. 

The decrease of $P_{\text{lin}}$, for the NV$^-$ centers, might be due to an increase in the local sample temperature, which could be unavoidable at high power excitation. The decrease of $P_{\text{lin}}$ at high powers is comparable to its decrease measured at a temperature of about 10~K. However, at high excitation powers we also observe an increase in $\theta$ only for the NV$^-$ ZPL, see Fig.~\ref{fig5}(d). Contrarily, an increase in $\theta$ is not observed at temperatures below 50~K. Since we can assume that the temperature increase due to high laser power excitation is far below such values, the increase in $\theta$ with rising laser power cannot be explained by heating effects only. One possible explanation is a change in the refractive indices for the left- and right-circularly polarized components of the ZPL emission induced by high-power exciting laser light.

\section{Conclusion} \label{conclsec}

We demonstrate that the zero phonon lines of both the negatively charged and neutral nitrogen vacancies as well as neutral vacancies are strongly linearly polarized, even at room temperature. The temperature dependences of the NV$^-$ ZPL indicate a direct phonon process between the excited states. Due to crystallographic properties of the NV centers, the angular dependence of their linear polarization degrees is anisotropic. This anisotropy not only characterizes the NV center orientation and crystal symmetry, but in turn also offers the possibility of selectively addressing linearly polarized states by a controlled change of the crystal orientation. Moreover, a magneto-optical rotation of the polarization plane of the NV$^-$ center emission can be seen at high external magnetic fields for low and room temperatures and the different crystal orientations studied. At high laser power and zero magnetic field, the rotation of the polarization plane is enhanced, which may be attributed to an optically induced change in the refractive indices of the right- and left-circular-polarized NV$^-$ ZPL components. In addition to that, the angular dependence of the linear polarization degree shows a phase shift in the (001) direction at strong magnetic fields, which may result from the magnetic torque of the applied field on the magnetic moments of the NV$^-$ center electrons.

We acknowledge financial support by the DFG in the frame of the ICRC TRR 160 (Project No. A1 and B2) and by the SB RAS (Project No. 0306-2016-0004).

\end{document}